\documentstyle[12pt]{article}

\newcommand{\be}{\begin{equation}}
\newcommand{\ee}{\end{equation}}
\newcommand{\bea}{\begin{eqnarray}}
\newcommand{\eea}{\end{eqnarray}}
\newcommand{\sptwo}{1.4}
\newcommand{\doublespace}{\edef\baselinestretch{\sptwo}\Large\normalsize}
\newcommand{\newsection}[1]{
\section{#1}
\setcounter{equation}{0}}
\renewcommand{\theequation}{\thesection.\arabic{equation}
}

\newcounter{newapp}
\renewcommand{\thenewapp}{\Alph{newapp}}

\begin{document}

\hspace*{\fill} PURD-TH-95-02 \\
\begin{center}
{\large\bf The Supercurrent In Supersymmetric Field
Theories}
\end{center}
{}~\\
\begin{center}
{\bf T.E. Clark and S.T. Love}\\
{\it Department of Physics\\
Purdue University\\
West Lafayette, IN 47907-1396}
{}~\\
{}~\\
\end{center}
\vspace{1.0in}
\begin{center}
{\bf Abstract}
\end{center}

A supercurrent superfield whose components include a conserved
energy-momentum tensor and supersymmetry current
as well as a (generally
broken) R-symmetry current is constructed for a generic effective
N=1 supersymmetric gauge theory. The general form of the R-symmetry
breaking is isolated. Included within the various special cases considered
is the identification of those models which exhibit an unbroken R-symmetry.
One such example corresponds to a non-linearly
realized gauge symmetry where the chiral field R-weight is required
to vanish.

\pagebreak

\doublespace

\newsection{Introduction}

The most general graded Lie algebra of symmetries of the S-matrix
of a relativistic quantum field theory is the direct product of
(extended) supersymmetry (SUSY) with some
internal symmetry \cite{HLS}. That is,
supersymmetry is the only possible extension of
the Poincar\'{e} space-time symmetries. Moreover, since models
possessing supersymmetry tend to exhibit a less singular
ultraviolet behavior than what would naively be expected,
one is naturally led to
explore the role of SUSY in possible extensions of general relativity
and quantum theories of gravity.  Indeed, supersymmetry plays a pivotal role
in many of the string theories \cite{GSW} which offer the promise of
incorporating gravity in a consistent quantum mechanical framework.

This softer ultraviolet behavior of supersymmetric theories
can be encoded in various non-renormalization theorems \cite{FL},
which, among other things, guarantees that
supersymmetric models are free of additive quadratic divergences even when
they contain fundamental scalar degrees of freedom.
This attribute
allows mass hiearchies which are established at tree level in such theories
to remain stable against quantum fluctuations and
has led to a considerable amount of activity in SUSY model
building \cite{IR}. Such a SUSY effective theory often arises as
the flat space-time limit of a some supergravity model which in turn can be
considered as the zero slope
limit of an underlying superstring theory.  The resulting SUSY
model will, in general, contain interaction terms beyond those appearing
in the perturbatively renormalizable case.

It has also been demonstrated \cite{Seib}\cite{SW} that the
restrictions imposed by SUSY may dictate that
certain exact results can be established
even after the inclusion of perturbative and nonperturbative
radiative corrections. A crucial
ingredient used in securing these results involves the (extended) SUSY
algebra \cite{WO} which, in turn, is related to the supersymmetry currents.
In addition, explorations \cite{POPITZ} continue into
the possibility of having
non-perturbative violations of the non-renormalization theorems and
dynamical supersymmetry breaking which in turn could provide for the
natural origin of the huge hierarchy between the Planck scale
and the electroweak scale. It has been argued that
the nature of the R-symmetry realization plays an
important role in determining the viability and calculability
of this potentiality.  Once again, the resultant SUSY models are generally
required to contain higher dimensional operators in order to secure a stable
ground state.

For perturbatively renormalizable models containing
Yang-Mills vector superfields and
(anti-) chiral superfields, it has been shown that
the supersymmetry current is intimately related to the energy-momentum
tensor and the R-symmetry current. In fact, a supercurrent
multiplet\cite{FZ} can be constructed such that its
components contain these currents. Futhermore, the
generalized (spinor) trace of the supercurrent not only describes the
(non-) conservation of these component currents, but also that
of the associated superconformal symmetry currents.  Since, at the present
time, many of the supersymmetric models being investigated
involve more general structures than those appearing in this
perturbatively renormalizable class,
we constuct, in this paper, the general form of the supercurrent
in a larger class of models
characterized by arbitrary superpotential and prepotential
functions as well as an arbitrary K\"ahler potential.

In the next section, we define the model action which is the general
supersymmetric and gauge invariant form containing at most two derivatives.
We also introduce functional differential operator representations for both the
internal gauge symmetry, which can be either linearly or non-linearly realized,
and the space-time Poincar\'{e}, supersymmetry and R- transformations.
Starting with the supercurrent trace identity, section
3 details the construction of the supercurrent which is secured by combining
the various space-time symmetries into a particular superfield structure with
the R-symmetry current as the lowest component.  So doing, we obtain the
general form of possible R-symmetry breaking.  As a special case, we
review the form of the supercurrent obtained in perturbatively renormalizable
SUSY models. In addition, we delineate the general criterion nessecary
for an unbroken R-symmetry.

\newpage

\newsection{The Supersymmetric and Gauge Invariant Action}

Through two derivatives, the most general supersymmetric
and gauge invariant action, $\Gamma$,
composed of Yang-Mills vector superfields, $V^A$ , and
matter (anti-) chiral superfields,
$(\bar\phi^{\bar i})~\phi^i$, which transform either linearly
or non-linearly under the gauge group G, is
\bea
\Gamma [\phi, \bar\phi, V] &=& \int dV K(\phi, \bar\phi , V)
+\int dS\left[ \frac{1}{2}
f_{AB}(\phi)W^{A\alpha}W^B_\alpha +P(\phi)\right]\cr
 & &\qquad\qquad +\int d\bar S\left[ \frac{1}{2} \bar
f_{AB}(\bar{\phi})
\bar W^{A}_{\dot\alpha}\bar W^{B\dot\alpha} +\bar
P(\bar\phi)\right] \, .
\label{action}
\eea
This action contains a locally invariant K\"ahler potential \cite{Z},
$K=K(\phi, \bar\phi , V),$
the SUSY Yang-Mills kinetic term multiplying (anti-) chiral
field dependent prepotential functions,
$(\bar f_{AB} (\bar\phi))$ $f_{AB} (\phi)$, and the (anti-)
chiral superpotential $(\bar P (\bar\phi))$ $P(\phi)$.  The
adjoint representation chiral spinor field strength $W_\alpha$ \cite{WB} is
\bea
W_\alpha & \equiv & W^A_\alpha t^A
=-\frac{1}{4}\bar D\bar D \left[e^{-2V}D_\alpha
e^{2V}\right] \, ,
\eea
where $t^A$ are the adjoint representation matrices,
$(t^A)_{BC} \equiv if_{BAC}$ and \\
$V\equiv t^A V^A$ is the matrix valued gauge field.  It
proves convenient to introduce the polynomial in $V$
combination
\be
\ell_{AB}\equiv \left( {e^{2V}-1\over V}\right)_{AB}
\ee
in terms of which we can write
\be
W_\alpha = -\frac{1}{4}\bar D\bar D \left[D_\alpha V^B
\ell_{BA} \right]t^A \, ,
\ee
which explicitly identifies the $W^A_\alpha$ spinors as
\bea
W^A_\alpha &=& -\frac{1}{4}\bar D\bar D \left[D_\alpha V^B \ell_{BA} \right]\,
{}.
\eea
Similarly the anti-chiral field strength is
\bea
\bar W_{\dot\alpha}&=& \bar W^A_{\dot\alpha}t^A= -
\frac{1}{4}DD
\left[e^{2V}\bar D_{\dot\alpha}e^{-2V}\right]\cr
 &=& -\frac{1}{4}DD
\left[ \bar D_{\dot\alpha}V^B\bar \ell_{BA}\right]t^A \, ,
\eea
where
\be
\bar \ell_{AB}=\left({e^{-2V}-1\over V}\right)_{AB}=-
\ell_{BA} \, ,
\ee
and
\be
\bar W^A_{\dot\alpha} =  -\frac{1}{4}DD \left[\bar
D_{\dot\alpha} V^B \bar \ell_{BA} \right] \, .
\ee

\subsection{Gauge Invariance}

The generators of the symmetry groups can be realized using
Ward identity functional differential operators acting on the
superfields.  The infinitesimal gauge transformations of the
fields are defined by the functional differential operator
\bea
\delta (\Lambda,\bar \Lambda) &\equiv & \int dS
\Lambda^AA^i_A (\phi){\delta\over\delta \phi^i} +  \int d\bar
S \bar\Lambda^A \bar A^{\bar i}_A
(\bar\phi){\delta\over\delta \bar\phi^{\bar i}}\cr
 & & \qquad -i\int dV \left(\Lambda^B \ell^{-1}_{BA}
+\bar\Lambda^B \bar \ell^{-1}_{BA} \right) {\delta\over
\delta V^A}\, ,
\eea
where $(\bar\Lambda^A)~\Lambda^A$ are the infinitesimal
(anti-) chiral superfields parameterizing  the gauge variation.
When applied directly to the fields themselves, this yields
their individual variations as
\bea
\delta (\Lambda,\bar \Lambda) \phi^i & = & \Lambda^AA
^i_A (\phi) \cr
\delta (\Lambda,\bar \Lambda) \bar\phi^{\bar i}&= &
\bar\Lambda^A \bar A^{\bar i}_A (\bar\phi) \cr
\delta (\Lambda,\bar \Lambda) V^A & = & -i\left[ \Lambda^B
\ell^{-1}_{BA} + \bar\Lambda^B \bar \ell^{-1}_{BA}\right] \, .
\label{gauge}
\eea
The (chiral) Killing vectors $A_A^i (\phi)$ and their (anti-chiral)
complex conjugates,
$\bar A_A^{\bar i} (\bar\phi)$, define the global
transformations of the matter fields, which are denoted by the
variation $\delta_A$, so that
\bea
\delta_A \phi^i &=& A_A^i (\phi)\cr
\delta_A \bar\phi^{\bar i} &=& \bar A_A^{\bar i} (\bar\phi) \, .
\eea
These (anti-) chiral Killing vectors obey their defining Lie
derivative or Killing equations
\bea
A^j_A A^i_{B},_{j} -A^j_B A^i_{A},_{j} &=& if_{ABC}
A^i_C \cr
\bar A^{\bar j}_A \bar A^{\bar i}_{B},_{\bar j} -\bar
A^{\bar j}_B \bar A^{\bar i}_{A},_{\bar j} &=& if_{ABC}
\bar A^{\bar i}_C \, ,
\label{lie}
\eea
where we have introduced a notation where subscripts
following commas denote differentiation so that,
for example, $A^i_{B},_{j}={\partial
A^i_B\over\partial\phi^j}$, $ \bar A^{\bar i}_{B},_{\bar j}=
{\partial \bar A^{\bar i}_B\over\partial\bar\phi^{\bar j}}$.
These equations are a direct consequence of the gauge transformation algebra
\be
\left[\delta (\Lambda,\bar \Lambda),~ \delta
(\Lambda^\prime,\bar \Lambda^\prime)\right] = i\delta
(\Lambda \times \Lambda^\prime,
{}~\bar \Lambda \times \bar\Lambda^\prime) \, ,
\label{alg}
\ee
where the cross product is defined by the totally antisymmetric
structure constant of the group, $f_{ABC}$, so that
$(\Lambda \times \Lambda^\prime )_A
=f_{ABC}\Lambda^B\Lambda^{\prime~C}$.

For linear realizations of the gauge symmetry, Eq.~(\ref{lie})
is solved by $A^i_A = i(T^A)_{ij} \phi^j$, where the
$T^A$ form a matrix representation (perhaps reducible) of
the group so that
 $[T^A, T^B ]=if_{ABC} T^C$.  On the other hand, for non-linear
realizations, such as in the case of supersymmetric
non-linear sigma models \cite{Z}\cite{susysigma}\cite{CL1}\cite{CL2},
the $A^i_A$ solving
Eq.~(\ref{lie}) and thus forming a realization of the algebra are
non-linear functions of the $\phi^i$.  Using the local $\phi^i$
transformation law, the superpotential, $P(\phi)$, is seen to
be locally invariant,
$\delta (\Lambda, \bar\Lambda) P(\phi)=\Lambda^A A_A^i
(\phi) P(\phi),_i =0$, provided it is globally invariant,
$\delta_A P(\phi) =A^i_A(\phi)  P(\phi),_i =0$.

The gauge group transformation of the Yang-Mills vector
superfields is defined via
\be
e^{2V^\prime}=e^{2(V+\delta V)}\equiv e^{-
i\Lambda}e^{2V} e^{+i\bar\Lambda} \, .
\label{trans}
\ee
For infinitesimal $\Lambda$ and $\bar\Lambda$, this
reduces, upon application of the
Baker-Campbell-Hausdorff formula, to
\be
\delta (\Lambda, \bar\Lambda)V^A
=\frac{1}{2}\left(\bar\Lambda^B
+\Lambda^B\right)f_{ABC} V^C
+\frac{i}{2}\left(\bar\Lambda^B -\Lambda^B\right) \left[
V\coth{V}\right]_{BA}\, ,
\ee
which can be shown to be identical to the last line of
Eq.~(\ref{gauge}).  Since this result is also
consistent with the group algebra (Eq.~(\ref{alg})),
this transformation also forms a
realization of the gauge group.
Using the vector field gauge transformation, it is readily
established that the field strength spinors transform as the
adjoint representation under gauge transformations:
\bea
\delta (\Lambda, \bar\Lambda) W_\alpha &=& i\left[
\Lambda, W_\alpha \right]  \cr
\delta (\Lambda, \bar\Lambda) \bar W_{\dot\alpha} &=&
i\left[ \bar\Lambda, \bar W_{\dot\alpha} \right] \, ,
\eea
or equivalently
\bea
\delta (\Lambda, \bar\Lambda) W_\alpha^A &=& i\left(
\Lambda^C t^C\right)_{AB} W_\alpha^B   \cr
\delta (\Lambda, \bar\Lambda) \bar W_{\dot\alpha}^A &=&
i\left( \bar\Lambda^C t^C\right)_{AB}
\bar W_{\dot\alpha}^B \, .
\eea
Since any non-trivial (anti-) chiral prepotential terms, $(\bar
f_{AB}(\bar\phi)) ~
f_{AB}(\phi)$ are constructed to transform as the product of the
(anti-) chiral adjoint representations of the gauge group so that,
\bea
\delta (\Lambda, \bar\Lambda) f_{AB} &=& i\left(
\Lambda^D t^D\right)_{AC}f_{CB}+i\left( \Lambda^D
t^D\right)_{BC} f_{AC}\cr
\delta (\Lambda, \bar\Lambda) \bar f_{AB} &=& i\left(
\bar\Lambda^D t^D\right)_{AC}\bar f_{CB}+i\left(
\bar\Lambda^D t^D\right)_{BC} \bar f_{AC}\, ,
\eea
it follows that the contractions
$W^{A\alpha}f_{AB}W^B_\alpha$ and $\bar
W^A_{\dot\alpha}
 \bar f_{AB} \bar W^{B\dot\alpha}$ are gauge invariant.

Mutatis mutandis, the vector field transformations define a
realization of the complexified chiral gauge group $G^{(+)}\times G^{(-)}$ with
superfield parameters
\be
\lambda^A_{\pm}\equiv \frac{1}{2}\left(\bar\Lambda^A \pm
\Lambda^A\right)
\ee
and corresponding gauge transformation functional
differential operators
\bea
\delta_A^{(+)} &=& \int dS \delta_A\phi^i{\delta\over
\delta\phi^i} + \int d\bar S \delta_A\bar\phi^{\bar
i}{\delta\over \delta\bar\phi^{\bar i}}
+\int dV f_{ABC}V^C {\delta\over \delta V^B} \cr
\delta_A^{(-)} &=& -\int dS \delta_A\phi^i{\delta\over
\delta\phi^i} + \int d\bar S \delta_A\bar\phi^{\bar
i}{\delta\over \delta\bar\phi^{\bar i}}
+\int dV i\left(V\coth{V}\right)_{AB} {\delta\over \delta
V^B} \, .\cr
 & &
\label{pmhat}
\eea
In terms of these variations,  the gauge transformations take
the form $\delta (\Lambda, \bar\Lambda)=
\lambda^A_+ \delta_A^{(+)} +\lambda^A_-
\delta_A^{(-)}$.  Moreover,  the $\delta_A^{(\pm)}$
variations obey the chiral algebra \cite{CL2} given by
\bea
\left[\delta^{(+)}_A ,\delta^{(+)}_B \right] &=&
f_{ABC}\delta_C^{(+)}\cr
\left[\delta^{(+)}_A ,\delta^{(-)}_B \right] &=&
f_{ABC}\delta_C^{(-)}\cr
\left[\delta^{(-)}_A ,\delta^{(-)}_B \right] &=&
f_{ABC}\delta_C^{(+)}\, .
\eea
As such, the Yang-Mills fields, $V^A$, transform in the
adjoint representation of the $G^{(+)}$ subgroup and provide
a non-linear realization of the $G^{(-)}$
subgroup.  Written in terms of the $\lambda_\pm$ superfield
parameters, the exponential transformation law of the Yang-
Mills fields is
\bea
e^{2V^\prime} &=& e^{-
i\Lambda}e^{2V}e^{+i\bar\Lambda}\cr
 &=& e^{-i(\lambda_+ -\lambda_- )}e^{2V} e^{i(\lambda_+
+\lambda_- )}\, ,
\eea
which in turn yields the infinitesimal transformation laws
\bea
\delta_A^{(+)}V^B &=& i(t^A)_{BC} V^C\cr
\delta_A^{(-)} V^B &=& i(V\coth{V})_{AB}\, ,
\eea
in agreement with Eq.~(\ref{pmhat}).

Using the $\delta^{(-)}_A$ variations for the matter
fields, the gauge invariant K\"ahler potential,
$K(\phi,\bar\phi, V)$, can be constructed \cite{CL1}\cite{CL2}
from the globally invariant K\"ahler
potential, $K_0 (\phi, \bar\phi)$.  To achieve this, we define
the pure chiral matter field transformation operators
$\delta_A^{(\phi\pm)}$ as
\bea
\delta_A^{(\phi +)} &=& \int dS \delta_A\phi^i{\delta\over
\delta\phi^i} + \int d\bar S \delta_A\bar\phi^{\bar
i}{\delta\over \delta\bar\phi^{\bar i}}\cr
\delta_A^{(\phi -)} &=& -\int dS \delta_A\phi^i{\delta\over
\delta\phi^i} + \int d\bar S \delta_A\bar\phi^{\bar
i}{\delta\over \delta\bar\phi^{\bar i}}\, .
\eea
Then using the commutation relation \cite{CL2}
\be
\left[ \delta (\Lambda, \bar\Lambda), e^{iV^A\delta_A^{(\phi -)}}
\right] = e^{iV^A\delta_A^{(\phi -)}} \left[ \lambda_-^B \left(
\tanh{\frac{1}{2}V}\right)_{BC} \delta_C^{(\phi -)} -\lambda_-
^B \delta_B^{(\phi -)}\right]\, ,
\ee
it follows that
\be
K(\phi, \bar\phi, V) \equiv e^{iV^A\delta_A^{(\phi -)}} K_0(\phi,
\bar\phi)
\ee
is locally gauge invariant
\be
\delta (\Lambda, \bar\Lambda) K(\phi, \bar\phi, V) = 0\, ,
\ee
provided $K_0 (\phi, \bar\phi)$ is globally invariant,
$\delta_A K_0 (\phi, \bar\phi) = 0$.
A globally invariant K\"ahler potential can always be found \cite{CL1}
when the group does not contain explicit $U(1)$
factors.  Moreover, in that globally noninvariant case, where $\delta_A
K_0(\phi, \bar\phi) =\bar F_A (\bar\phi) +F_A (\phi) \neq 0$,
a locally invariant form can also be constructed \cite{CL1}.

\subsection{Supersymmetry}

In addition to its gauge invariance, the action, Eq.~(\ref{action}), is
also supersymmetric and Poincar\'{e} invariant.  These global
superspace symmetries are represented by superspace
differential operators on the superfields, which in turn can be
used to construct functional differential operators
representing the generators of the symmetries.  The
supersymmetry transformations are given by
\bea
\delta^Q_\alpha \Phi (x, \theta, \bar\theta) &=& \left[
{\partial\over \partial\theta} +i\sigma^\mu \bar\theta
\partial_\mu \right]_\alpha \Phi (x, \theta, \bar\theta)\cr
\delta^{\bar Q}_{\dot\alpha} \Phi (x, \theta, \bar\theta) &=&
\left[ -{\partial\over \partial\bar\theta} -i\theta\sigma^\mu
\partial_\mu \right]_{\dot\alpha} \Phi (x, \theta, \bar\theta) \, ,
\eea
where $\Phi$ is any of the superfields $\phi,~\bar\phi$ or
$V$, while the variation of the fields under  space-time
translations is given by
\be
\delta^P_\mu \Phi (x, \theta, \bar\theta) = \partial_\mu \Phi (x,
\theta, \bar\theta)\, .
\ee
These variations can be combined to form the Ward identity
functional differential operator representing the  generators of
the symmetries.  The functional differential operators
corresponding to the supersymmetry charges $Q_\alpha$ and
$\bar Q_{\dot\alpha}$ are
\bea
\delta^Q_\alpha &=& \int dS \delta^Q_\alpha \phi^i
{\delta\over\delta\phi^i} +\int d\bar S \delta^Q_\alpha \bar\phi^{\bar{i}}
{\delta\over\delta\bar\phi^{\bar{i}}} +\int dV \delta^Q_\alpha V
{\delta\over\delta V}\cr
\delta^{\bar Q}_{\dot\alpha} &=&  \int dS \delta^{\bar
Q}_{\dot\alpha} \phi^i {\delta\over\delta\phi^i} +\int d\bar S
\delta^{\bar Q}_{\dot\alpha} \bar\phi^{\bar{i}}
{\delta\over\delta\bar\phi^{\bar{i}}}+\int dV \delta^{\bar
Q}_{\dot\alpha} V {\delta\over\delta V}\, ,
\eea
while those corresponding to the space-time translation
generators $P^\mu$ are
\bea
\delta^P_\mu &=& \int dS \delta^P_\mu \phi^i
{\delta\over\delta\phi^i} +\int d\bar S \delta^P_\mu \bar\phi^{\bar{i}}
{\delta\over\delta\bar\phi^{\bar{i}}}+\int dV \delta^P_\mu V
{\delta\over\delta V}\, .
\eea
Similar expressions also hold for Lorentz transformations.
These variations satisfy an algebra analogous
to the one satisfied by the global
symmetry generators. For example, while the supersymmetry
charges anti-commute to yield the momentum operator,
\be
\left\{ Q_\alpha ,\bar
Q_{\dot\alpha} \right\} = 2\sigma^\mu_{\alpha\dot\alpha}P_\mu  \, ,
\ee
it is readily seen that
\be
\left\{ \delta^Q_\alpha, \delta^{\bar Q}_{\dot\alpha} \right\}
= -2i\sigma^\mu_{\alpha\dot\alpha} \delta^P_\mu\, .
\ee
By construction, the action
$\Gamma$ is invariant under supersymmetry and space-time
translation
transformations so that
\bea
\delta^Q_\alpha ~\Gamma [ \phi, \bar\phi, V ] &=&0\cr
\delta^{\bar Q}_{\dot\alpha}~ \Gamma [ \phi, \bar\phi, V ]
&=&0\cr
\delta^P_\mu ~\Gamma [ \phi, \bar\phi, V ] &=&0\, .
\eea

The action may also be invariant under R-symmetry or
some global internal symmetries.  In particular, the generator
of R-symmetry transformations is given by
\bea
\delta^R &=& \int dS \delta^R \phi^i {\delta\over\delta\phi^i}
+\int d\bar S \delta^R \bar\phi^{\bar{i}} {\delta\over\delta\bar\phi^{\bar{i}}}
+\int
dV \delta^R V {\delta\over\delta V}\, ,
\eea
where the explicit R-symmetry transformations of the fields
are defined by
\be
\delta^R \Phi (x, \theta, \bar\theta )= i\left[ n_\Phi + \theta^\alpha
{\partial\over\partial\theta^\alpha} + \bar\theta_{\dot{\alpha}}
{\partial\over\partial\bar\theta_{\dot{\alpha}}}\right] \Phi (x, \theta,
\bar\theta)\, ,
\ee
with $n_\Phi$  the R-weight of the superfield $\Phi$.  Since
the vector superfield is real, its R-weight must be zero: $n_V
=0$.  In general, the R-weight of the chiral superfields, $n_\phi$,
is arbitrary.  In some cases, however, it can be fixed so as to
make the superpotential R-invariant.  Moreover, as shown in
the Appendix, if the chiral superfield transforms non-linearly
under the gauge group, its R-weight must be zero: $n_\phi =
0$.  The Weyl spinor supersymmetry charges $Q_\alpha$ and
$\bar Q_{\dot\alpha}$ form a representation of the chiral R-
symmetry given by
\bea
\left[R, Q_\alpha \right] &=& Q_\alpha \cr
\left[R, \bar Q_{\dot\alpha} \right] &=& -\bar
Q_{\dot\alpha}\, .
\eea
Likewise, it follows that the Ward identity functional
differential operators obey the analogous algebra
\bea
\left[ \delta^R , \delta^Q_\alpha \right] &=& -i
\delta^Q_\alpha \cr
\left[ \delta^R , \bar\delta^{\bar Q}_{\dot\alpha} \right] &=&
i \bar\delta^{\bar Q}_{\dot\alpha}\, .
\eea

Using Noether's theorem, the (non-) conserved currents
corresponding to these transformations can be constructed
from the action.  Since $\Gamma$ is supersymmetric and
translation invariant, the corresponding supersymmetry
currents, $Q^\mu_\alpha$, $\bar Q^\mu_{\dot\alpha}$, and
the energy-momentum tensor, $T^{\mu\nu}$, are conserved and satisfy
\bea
\partial_\mu Q^\mu_\alpha(x) &=& \delta^Q_\alpha (x)
{}~\Gamma\cr
\partial_\mu \bar Q^\mu_{\dot\alpha}(x) &=& \bar\delta^{\bar
Q}_{\dot\alpha} (x) ~\Gamma\cr
\partial_\mu T^{\mu\nu}(x) &=& \delta^{P\nu} (x)~\Gamma \, .
\label{cc}
\eea
Here $\delta^Q_\alpha (x)$, $\bar\delta^{\bar
Q}_{\dot\alpha} (x)$ and $\delta^P_\mu (x)$ are the local
SUSY and translation functional differential operators
respectively.  The corresponding global transformation
functional differential operators, $\delta^Q_\alpha,
\bar{\delta}^{\bar{Q}}_{\dot{\alpha}}, \delta^P_\mu$,
are constructed by integrating
the local operators over space-time.  Thus, for example,
$\delta^Q_\alpha =\int d^4 x
\delta^Q_\alpha (x)$, is the global SUSY variation.  It follows that
the currents of Eq.~(\ref{cc}) can be modified (improved) by the
addition of Belinfante terms or total space-time divergences
of Euler derivatives of the action (contact terms) without alterring the
form of the current conservation law or the time independent charges.

Similarly, the R-current can
be constructed via Noether's theorem as
\be
\partial_\mu R^\mu(x) =\delta^R (x) ~\Gamma -i S_R(x)\, ,
\ee
where $S_R(x)$ describes the explicit R-symmetry breaking
of the action.  Integrating this equation over space-time gives
\be
\delta^R ~\Gamma[\phi, \bar{\phi}, V] = i \int d^4 x S_R(x) \, ,
\ee
which constitutes the global R Ward identity.

The R-current so defined can be extended so as to form an entire
superfield with
$R^\mu (x)$ as its lowest component.
It is this multiplet with appropriately defined improved supersymmetry
currents and energy-momentum tensor which constitutes the
supercurrent.  By construction the supercurrent contains a
(non-) conserved R-symmetry current $R^\mu(x)$ as the
lowest component with conserved supersymmetry currents,
$Q^\mu_\alpha(x)$, $\bar Q^\mu_{\dot\alpha}(x)$,
and symmetric energy-momentum tensor, $T^{\mu\nu}(x)$ , in
higher components \cite{FZ}\cite{CPS}\cite{PS1}\cite{screfs}.

\pagebreak

\newsection{The Supercurrent}

In general, the generators of global symmetry transformations
can be obtained from the symmetry currents using Noether's
theorem.  For the superconformal symmetries, all
superconformal currents can be gleaned from the
supercurrent \cite{CPS}\cite{PS1}.  The supercurrent is just
the superfield of currents whose first component is given by
the R-symmetry current and which, moreover, contains the
supersymmetry currents and the improved energy-momentum tensor as
higher dimension components.  It has been shown on very
general grounds \cite{CPS}\cite{PS1} that the real supercurrent,
$V_{\alpha\dot\alpha}=\frac{1}{2}\sigma^\mu_{\alpha\dot\alpha}
V_\mu$, must satisfy a general set of spinor derivative (trace)
equations of the form
\bea
\bar D^{\dot\alpha} V_{\alpha\dot\alpha} &=& -
2\hat\delta_\alpha \Gamma +B_\alpha -2D_\alpha S\cr
D^\alpha V_{\alpha\dot\alpha} &=& -
2\hat{\bar\delta}_{\dot\alpha} \Gamma +\bar B_{\dot\alpha} -2\bar
D_{\dot\alpha} \bar S \, .
\label{trace}
\eea
The $B_\alpha$ and $\bar B_{\dot\alpha}$ are restricted to
obey $D^\alpha B_\alpha =\bar D_{\dot\alpha} \bar
B^{\dot\alpha}$ while ($\bar S$) $S$ is a (anti-) chiral
superfield, ($D^\alpha \bar S =0$) $\bar D_{\dot\alpha} S
=0$.  In order for $V_{\alpha\dot\alpha}$  to contain a
conserved
energy-momentum tensor $T^{\mu\nu}$, and supersymmetry
currents $Q^\mu_\alpha$ and $\bar Q^\mu_{\dot\alpha}$, it
must be that the symmetry breaking terms ($\bar
B_{\dot\alpha}$)  $B_\alpha$ and ($\bar S$) $S$ cannot both
be non-zero simultaneously. The local superspace Ward
identity functional differential operators, $\hat\delta_\alpha$,
$\hat{\bar\delta}_{\dot\alpha}$, are defined as
\bea
\hat{\bar\delta}_{\dot\alpha} &\equiv& n_\phi\bar D_{\dot\alpha}
\left( \bar\phi^{\bar i}{\delta\over \delta\bar\phi^{\bar
i}}\right)
+2\left(\bar D_{\dot\alpha}\bar\phi^{\bar i}\right)
{\delta\over \delta \bar\phi^{\bar i}}\cr
 & &-2\left( DD\bar D_{\dot\alpha}V^A\right)
{\delta\over \delta V^A}+2\left( \bar
D_{\dot\alpha}V^A\right)
DD{\delta\over \delta V^A} \cr
 & & +2D^\alpha\left[\bar D_{\dot\alpha} V^D D_\alpha
V^A
\left(\ell_{AB},_D\ell^{-1}_{BC}+\bar \ell_{DB},_A
\bar \ell^{-1}_{BC}\right) {\delta\over \delta
V^C}\right]\cr
\hat\delta_{\alpha} &\equiv& n_\phi D_{\alpha}
\left( \phi^{ i}{\delta\over \delta\phi^{ i}}\right)
+2\left( D_{\alpha}\phi^{i}\right)
{\delta\over \delta \phi^{i}}\cr
 & &-2\left( \bar D\bar D D_{\alpha}V^A\right)
{\delta\over \delta V^A}+2\left( D_{\alpha}V^A\right)
\bar D\bar D{\delta\over \delta V^A} \cr
 & & +2\bar D_{\dot\alpha}\left[D_{\alpha} V^D \bar
D^{\dot\alpha} V^A
\left(\ell_{AB},_D\ell^{-1}_{BC}+\bar \ell_{DB},_A
\bar \ell^{-1}_{BC}\right) {\delta\over \delta
V^C}\right]\, .
\eea
Note that when restricted to Abelian gauge fields, the last
lines on the right hand side of each equation  vanishes.  The form of these
variations is such that all the superconformal transformations
can be secured by acting on them with appropriate spinor
derivatives and then constructing their various space-time
moments.  In particular, defining the local variation
\be
\hat\delta \equiv i( D^\alpha \hat\delta_\alpha -\bar D_{\dot\alpha}
\hat{\bar\delta}^{\dot\alpha})\, ,
\label{R}
\ee
then its space-time integral
\be
\delta =\int d^4 x \hat\delta   \, ,
\ee
forms the superfield containing the (previously defined)
R symmetry, supersymmetry
and space-time translation functional differential operators, $
\delta^R$, $\delta_\alpha^Q$, $\bar\delta^{\bar
Q}_{\dot\alpha}$ and $ \delta_\mu^P$, as
\be
\delta = \delta^R -i\theta^\alpha \delta_\alpha^Q +
i\bar\theta_{\dot\alpha}\bar\delta^{\bar Q\dot\alpha} -
2\theta\sigma^\mu \bar\theta \delta_\mu^P \, .
\ee
Note that alternate forms for
$\hat\delta_\alpha$ and $\hat{\bar\delta}_{\dot\alpha}$ can also be
defined by adding various terms which take the form of
additional total derivatives of contact terms (improvements)
or have the effect of
changing the relation of these variations to the conformal
transformations \cite{PS1}. The conservation of the
supersymmetry currents, $Q^\mu_\alpha,
\bar Q^\mu_{\dot\alpha}$, and the energy-momentum tensor,
$T^{\mu\nu}$, Eq.~(\ref{cc}),
follows from Eq.~(\ref{trace}) provided either (or both) $S$ or $B$ to vanish,
which alternative being a model dependent question \cite{CPS}\cite{PS1}.

Applying the spinor derivative construction of Eq.~(\ref{R})
to Eq.~(\ref{trace}) yields the space-time divergence
equation for the supercurrent
\bea
\partial^\mu V_\mu &=& \frac{1}{2i}\left\{ D^\alpha , \bar
D^{\dot\alpha} \right\} V_{\alpha\dot\alpha}\cr
 &=& \delta \Gamma -i\left( \bar D\bar D\bar S -
DDS\right)\, .
\eea
The $\theta$, $\bar\theta$ independent component of this
equation gives the R-current Ward identity
\be
\partial^\mu R_\mu =  \delta^R (x) \Gamma -i\left( \bar
D\bar D\bar S -DDS\right)\vert_{\theta =\bar\theta = 0}\, .
\label{Rcurrent}
\ee
If $S\neq 0$, the $R$-symmetry is explicitly broken.  Note
that in such a case, in order for the supercurrent to contain a
conserved supersymmetry current and energy-momentum
tensor, it is required that $B=0$.  The construction of all the
superconformal currents along with their associated Ward identities
and anomalies is detailed in references \cite{CPS}\cite{PS1}.
Besides the R-symmetry current
constructed above as the $\theta$, $\bar\theta$ independent
component of the supercurrent itself,
\be
R_\mu =V_\mu|_{\theta=\bar{\theta}=0}  \, ,
\ee
the supersymmetry
currents and the energy-momentum tensor can similarly be constructed as
the $\theta$, $\bar\theta$ independent components of certain combinations
of spinor derivatives acting on the supercurrent as:
\bea
Q_{\mu\alpha} &=& i\left(D_\alpha V_\mu -\left(\sigma_\mu
\bar\sigma^\nu D\right)_\alpha V_\nu \right)|_{\theta=\bar{\theta}=0}\cr
\bar Q_{\mu\dot\alpha} &=& -i\left(\bar D_{\dot\alpha}
V_\mu -\left(\bar\sigma_\mu\sigma^\nu \bar D
\right)_{\dot\alpha} V_\nu \right)|_{\theta=\bar{\theta}=0}\cr
T_{\mu\nu} &=& -
\frac{1}{16}\left(V_{\mu\nu}+V_{\nu\mu} -
2g_{\mu\nu}V_\rho^\rho\right)|_{\theta=\bar{\theta}=0} \, ,
\label{susyem}
\eea
where the superfield $V_{\mu\nu}$ is defined as
\be
V_{\mu\nu} = \left( D\sigma _\mu \bar D -\bar
D\bar\sigma_\mu D\right) V_\nu \, .
\ee
The remaining superconformal currents and angular
momentum tensor can be constructed as space-time
moments of Eq.~(\ref{susyem}). For example the
dilatation current is given by $D_\mu =x^\nu T_{\mu\nu}$.

Given an action $\Gamma$ and the variations
$\hat\delta_\alpha$ and $\hat{\bar\delta}_{\dot\alpha}$,
the $V_{\alpha\dot\alpha}$, $B_\alpha$, $\bar
B_{\dot\alpha}$, $S$ and $\bar S$ are constructed so as to satisfy the trace
equations (\ref{trace}).  Towards this end, it is necessary
to use the field equations for the matter and Yang-Mills
superfields.  Functionally differentiating the action of Eq.~(\ref{action})
with respect to the chiral and anti-chiral matter
fields yields
\bea
{\delta \Gamma\over \delta \phi^i} &=& -\frac{1}{4}\bar
D\bar D
K,_i +P,_i -2 f_{AB},_i W^AW^B\cr
{\delta \Gamma\over \delta \bar\phi^{\bar i}} &=& -
\frac{1}{4}DD
K,_{\bar i} +\bar P,_{\bar i} -2 \bar f_{AB},_{\bar i}
\bar W^A \bar W^B .
\label{matter}
\eea
A useful form of the
Yang-Mills field equations \cite{FP} is obtained by
introducing the gauge covariant spinor derivatives ${\cal
D}_\alpha$ and $\bar{\cal D}_{\dot\alpha}$ for the chiral
field strength spinors as,
\bea
{\cal D}^\alpha W_\beta &\equiv & e^{-2V}D^\alpha \left[
e^{2V}W_\beta e^{-2V} \right] e^{2V} \cr
 &=&  D^\alpha W_\beta +\Omega^\alpha W_\beta +W_\beta
\Omega^\alpha\cr
\bar{\cal D}_{\dot\alpha} \bar W^{\dot\beta} &\equiv &
e^{2V}\bar D_{\dot\alpha}
 \left[ e^{-2V}\bar W^{\dot\beta} e^{2V} \right] e^{-2V}\cr
&=&  \bar D_{\dot\alpha}\dot W^{\dot\beta}
+\bar\Omega_{\dot\alpha} \bar W^{\dot\beta} +
\bar W^{\dot\beta} \bar\Omega_{\dot\alpha} \, ,
\eea
with
\bea
\Omega_\alpha &\equiv& e^{-2V}D_\alpha e^{2V} =
\left(D_{\alpha} V^A \ell_{AB} \right) t^B \equiv \Omega^B_\alpha t^B\cr
\bar \Omega_{\dot\alpha} &\equiv& e^{2V}D_{\dot\alpha}
e^{-2V} = \left(\bar D_{\dot\alpha} V^A \bar\ell_{AB}
\right) t^B \equiv \bar{\Omega}^B_{\dot{\alpha}} t^B \, .
\eea
Alternatively, these covariant derivatives can be written as
\bea
\left({\cal D}^\alpha W_{\beta}\right)^C &=& \left( e^{-
2V}\right)_{CB} D^{\alpha} \left[
\left(e^{2V}\right)_{BA}W^A_{\beta}\right]\cr
 & & \cr
 &=& D^{\alpha} W^C_{\beta} + if_{CBA} \Omega^\alpha_B W^A_{\beta} \cr
 & & \cr
\left(\bar{\cal D}_{\dot\alpha}\bar W^{\dot\beta}\right)^C
&=& \left( e^{2V}\right)_{CB}
\bar D_{\dot\alpha} \left[ \left(e^{-2V}\right)_{BA}\bar
W^{A\dot\beta}\right]\cr
 & & \cr
 &=& \bar D_{\dot\alpha}\bar W^{C\dot\beta} +
if_{CBA}\bar{\Omega}^B_{\dot{\alpha}} \bar W^{A\dot\beta} \, .
\eea
The field equations for the gauge fields can then be cast as
\bea
{\delta\Gamma\over \delta V^A} &=& -\ell_{AB}
\left( {\cal D}^\alpha F_\alpha\right)^B -\bar \ell_{AB}
\left( \bar{\cal D}_{\dot\alpha}  \bar F^{\dot\alpha}\right)^B
\cr
 &=&  -\ell_{AB}
\left( {\cal D}^\alpha F_\alpha\right)^B +
\left( \bar{\cal D}_{\dot\alpha}  \bar F^{\dot\alpha}\right)^B
\ell_{BA} \, ,
\label{vector}
\eea
where we have introduced the auxiliary field strength spinors
$F_\alpha$ and $\bar F_{\dot\alpha}$ defined as
$F^\alpha \equiv t^A f_{AB}W^B_\alpha$ and $\bar
F_{\dot\alpha} \equiv t^A \bar f_{AB} \bar
W^B_{\dot\alpha}$.
In addition to these dynamical relations, the field strength
spinors also satisfy the Bianchi identities
\be
{\cal D}^\alpha W_\alpha =  -e^{-2V} \left( \bar{\cal
D}_{\dot\alpha} \bar W^{\dot\alpha}\right)e^{2V} \, ,
\ee
which can alternatively be written as
\be
\left( \bar{\cal D}_{\dot\alpha} \bar W^{\dot\alpha}\right)^A
=
-\left( e^{2V}\right)_{AB}
\left( {\cal D}^\alpha W_\alpha \right)^B \, ,
\ee
or in further detail
\be
\bar D_{\dot\alpha} \left[ \left(e^{-2V}\right)_{AB} \bar
W^{B\dot\alpha}\right] = - \left(e^{-
2V}\right)_{AC}D^\alpha\left[\left(e^{2V}\right)_{CB}
W^B_\alpha\right] \, .
\label{bianchi}
\ee

Application of the Ward identity operator
$\hat{\bar\delta}_{\dot \alpha}$ to the general action of
Eq.~(\ref{action}) and
exploiting the field equations (\ref{matter}), (\ref{vector})
along with the
Bianchi identity (\ref{bianchi}), the supercurrent trace
equation (\ref{trace}) is seen to be
satisfied with the supercurrent identified as
\bea
V_{\alpha\dot\alpha} &=& 16\left[ \bar
W_{\dot\alpha}^A (e^{2V})_{AB}f_{BC}W_\alpha^C
- W_\alpha^A (e^{-2V})_{AB}{\bar f}_{BC}{\bar W}_{\dot\alpha}^C\right]\cr
 & & -\frac{2}{3}\left[ D_\alpha ,\bar D_{\dot\alpha}\right]
K
+2K,_{i\bar i} {\cal
D}_\alpha \phi^i  \bar{\cal D}_{\dot\alpha} \bar\phi^{\bar i} \, .
\eea
Here we have introduced the gauge covariant spinor
derivatives for the matter fields defined as
\bea
{\cal D}_\alpha \phi^i &\equiv & D_\alpha \phi^i -
i\Omega^B_\alpha A^i_B(\phi)\\
 &=& D_\alpha \phi^i -iD_\alpha V^A \ell_{AB} A^i_B
(\phi) \cr
\bar{\cal D}_{\dot\alpha} \bar\phi^{\bar i} &\equiv & \bar
D_{\dot\alpha} \bar\phi^{\bar i} -i\bar{\Omega}^B_{\dot{\alpha}}\bar{A}
^{\bar{i}}_B (\bar{\phi})\\
 &=& \bar D_{\dot\alpha} \bar\phi^{\bar i} -i\bar
D_{\dot\alpha} V^A \bar\ell_{AB}
\bar A^{\bar i}_B (\bar\phi) \, ,
\eea
which have the gauge variations
\bea
\delta \left(\Lambda , \bar\Lambda \right) {\cal D}_\alpha
\phi^i &=& \Lambda^A A^i_A,_j (\phi) {\cal D}_\alpha
\phi^j \cr
\delta \left(\Lambda , \bar\Lambda \right) \bar{\cal
D}_{\dot\alpha} \bar\phi^{\bar i} &=& \bar\Lambda^A \bar
A^{\bar i}_A,_{\bar j} (\bar\phi) \bar{\cal D}_{\dot\alpha}
\bar\phi^{\bar j} \, .
\eea
Note that $V_{\alpha\dot\alpha}$ is manifestly real and
gauge invariant.
In addition, one finds explicitly that $\bar B_{\dot\alpha}=0$, while
the anti-chiral breaking terms have the form
\bea
\bar S &=& -\frac{1}{4}DD\left( -\frac{2}{3} K -n_{\phi}\left(
\bar\phi^{\bar i} K,_{\bar i}\right)\right)
+8n_{\phi}\bar\phi^{\bar i}\bar W^A_{\dot\beta}\bar f_{AB},_{\bar
i}\bar
W^{B\dot\beta} \cr
 & &-\left( 2\bar P +n_{\phi} \bar\phi^{\bar i}\bar P,_{\bar i}\right) \, .
\eea
Since $\bar S \neq 0$, the R symmetry is, in general, explicitly broken.

Using this general form of the supercurrent and its associated
Ward identity,
various special cases can be considered.  First of all, the
form for $V_{\alpha\dot\alpha}$ and $\bar S$ reduce to
their previously established values \cite{PS1},
\bea
V_{\alpha\dot\alpha} &= & 32\bar W_{\dot\alpha} e^{2V^A
t^A}W_\alpha -\frac{2}{3}
\left[ D_\alpha ,\bar D_{\dot\alpha}\right] (\phi\, e^{2V^A
T^A}\bar\phi)
+2{\cal D}_\alpha \phi\, e^{2V^A T^A}\bar{\cal
D}_{\dot\alpha} \bar\phi \cr
\bar S  & = & \frac{(2+3n_{\phi})}{12}DD  (\phi\, e^{2V^A
T^A}\bar\phi)
-(2\bar P +n_{\phi}\bar\phi^{\bar i} \bar P,_{\bar i})  \, ,
\eea
when the model is restricted to be renormalizable so that $K=\phi \,
e^{2V\cdot T}\bar\phi$ and $P$ is at most trilinear in $\phi$
while $f_{AB}=\delta_{AB}$.  For conformal R-weight $n_\phi =-
\frac{2}{3}$ and no mass or linear terms in $P$, the
breaking terms vanish, $S=0$, and the R-current is
conserved.

For the particular case when the gauge symmetry is non-
linearly
realized on the chiral matter fields with a vanishing
superpotential,
it follows (see Appendix) that the $\phi$ field R-weight must
be zero:
$n_\phi = 0$. The supercurrent can then be cast as
\be
V_{\alpha\dot\alpha}=  16\left[ \bar
W_{\dot\alpha}^A(e^{2V})_{AB}f_{BC}W_\alpha^C
- W_\alpha^A (e^{-2V})_{AB}{\bar f}_{BC}{\bar W}_{\dot\alpha}^C\right]
+2{\cal{D}}_\alpha\phi^i K,_{i\bar i}\bar {\cal{D}}_{\dot
\alpha} \bar \phi^{\bar i} \, ,
\ee
with $\bar S =0$ and $\bar B_{\dot \alpha}= DD\bar D_{\dot
\alpha} K$.
Since $\bar S=0$, this form of the supercurrent
not only leads to a conserved supersymmetry current and
energy-momentum tensor, but also to a conserved R-
symmetry current.

When the chiral matter fields form a linear representation of
the
gauge group, the axial R-weight $n_\phi$ is
arbitrary.  If, however, the superpotential and prepotential are R-invariant so
that
\bea
2\bar P +n_\phi\bar\phi^{\bar i}\bar P,_{\bar i}&=&0,\\
\bar f_{AB}=f_{AB}&=&\delta_{AB} \, ,
\eea
while the gauged K\"ahler potential possesses an additional global,
axial $U_A (1)$
symmetry, so that
\be
\phi^i K,_{i} -\bar\phi^{\bar i}K,_{\bar i}=0 \, ,
\ee
then the $\bar S$ breaking term can again be traded for a $B$
breaking with a suitable modification of the supercurrent. So
doing, we
find

\bea
V_{\alpha\dot\alpha} &=& 32 \bar
W_{\dot\alpha}e^{2V}W_\alpha \cr
 & & +2 {\cal{D}}_\alpha \phi^i K,_{i\bar i}\bar {\cal{D}}_
{\dot \alpha}\bar \phi^{\bar \phi}+\frac{n_\phi}{2}\left[ D_\alpha
,\bar
D_{\dot\alpha}\right] \left( \phi^i K,_{i} +\bar\phi^{\bar
i}K,_{\bar i}\right)\cr
\bar S &=& 0\cr
\bar B_{\dot\alpha} &=& -\frac{1}{4}DD\bar D_{\dot\alpha}
\left[ -4K
-3n_\phi \left( \phi^i K,_{i} +\bar\phi^{\bar i}K,_{\bar
i}\right)\right]\, .
\eea

The quantization of the gauge models
requires the introduction of gauge fixing and Fadeev-Popov
terms to the Lagrangian \cite{PS1}\cite{FP}.  With their inclusion,
the action ceases to be gauge invariant but becomes
BRS invariant.  A detailed account of the supercurrent
construction in renormalizable models with BRS invariance can be
found in the literature \cite{PS1}.  When the
quantum corrections are taken into account, the
divergence of the
R-current, the $\gamma$-trace of the supersymmetry current
and the Lorentz trace of the
energy-momentum tensor are anomalous with the
renormalization group $\beta$ function as the anomaly
coefficient \cite{FZ}\cite{CPS}\cite{GGRS}.  The nature of
these radiative corrections for the renormalizable N=1 SUSY
models has been investigated and
reviewed \cite{PS1}\cite{GMZ}\cite{SV}\cite{K}.  For certain gauge models, the
$\beta$ function has been shown \cite{Seib}\cite{SW} or argued \cite{S2}
to vanish identically at a superconformal fixed point. This, in turn, fixes
the R-weights.

In all of the above, discussion has been restricted to the case of
linearly realized supersymmetry, while the gauge symmetry was allowed to
be realized either lineraly or non-linearly. For completeness, let us recall
the case of a non-linearly realized supersymmetry. Indeed in the absence of
explicit breakings, if supersymmetry is to be realized in nature, it must be
as a spontaneously broken symmetry. At high energy, the short distance
behavior of the theory will be unaffected by the soft spontaneous SUSY
breaking of the ground state. The structure of the supercurrent will
be identical to the unbroken case.  At low energy, the spontaneously
broken SUSY can be described by the Akulov-Volkov effective
Lagrangian \cite{AV}.  For this model, a supercurrent has also been
constructed \cite{CL3}. It again contains the conserved R-current as the
lowest component and conserved
supersymmetry current and the energy-monentum tensor as higher components.
In this case, the R-current is simply given by $R^\mu = -2\kappa^2 \lambda
\sigma_\nu \bar{\lambda} T^{\nu\mu}$,
where $\lambda$ is the Goldstino field and $\kappa$ is its decay
constant and $T^{\mu\nu}$ is the improved energy-momentum tensor.

\smallskip

This work was supported in part by the U.S. Department
of Energy under grant DE-AC02-76ER01428 (Task B).
\pagebreak

\setcounter{equation}{0}
\renewcommand{\theequation}{\thenewapp.\arabic{equation}
}

\section*{Appendix A }

In this appendix, we show that the R-weight, $n_\phi$, of any
chiral superfield transforming non-linearly under an internal
symmetry transformation must vanish: $n_\phi = 0$.  This
demonstration employs the algebra
\bea
\left[ T_A, T_B \right] &=& if_{ABC} T_C\cr
\left[ R, T_A\right] &=& 0\, ,
\eea
along with the chiral field transformation laws
\bea
\delta_A \phi^i &=& \frac{1}{i} \left[ T_A ,\phi^i \right] =
A_A^i (\phi) \cr
\delta^R \phi^i &=& \frac{1}{i}\left[ R, \phi^i \right] = i \left(
n_\phi +\theta^\alpha{\partial\over\partial\theta^\alpha}
+\bar\theta_{\dot{\alpha}}{\partial\over\partial\bar\theta_{\dot{\alpha}}}
\right) \phi^i  \, .
\eea
Using these relations, the Jacobi identity
\be
0=\left[\left[R,T_A\right],\phi^i\right] + \left[\left[\phi^i,
R\right], T_A \right] + \left[\left[T_A,\phi^i\right],R\right] \, ,
\ee
reduces to
\be
0= n_\phi \left( A^i_A -A^i_{A},_{j}\phi^j\right) \, .
\ee
For non-linear realizations
\be
A^i_{A},_{j}\phi^j \neq A^i_A \, ,
\ee
and hence we conclude that $n_\phi = 0$.

\pagebreak

\newpage
\end{document}